%% file: paper.tex
\documentclass[runningheads]{llncs}
\usepackage[T1]{fontenc}
\usepackage{graphicx}
\usepackage{xspace}
\usepackage{hyperref}
\usepackage{color}

\usepackage{ifthen}
\usepackage{amsfonts} 
\DeclareMathSymbol{\shortminus}{\mathbin}{AMSa}{"39}

\usepackage{mwe} %
\usepackage{color}
\usepackage[svgnames,table]{xcolor}
\usepackage{amsmath} %
\usepackage{amssymb}
\usepackage{booktabs}
\usepackage{multirow}
\usepackage{placeins}

\usepackage{caption}
\usepackage{subcaption}
\usepackage{textcomp} 

\begin{document}

\title{Robust vertebra identification using simultaneous node and edge predicting Graph Neural Networks}
\titlerunning{Robust vertebra identification using simult. node and edge predicting GNNs}
\author{{Vincent Bürgin
\inst{1,2\,}%
\thanks{This work is supported by the DAAD program Konrad Zuse Schools of Excellence in Artificial Intelligence, sponsored by the Federal Ministry of Education and Research}%
}
\and
{Raphael Prevost\inst{1}
}
\and~{Marijn F. Stollenga\inst{1}
}}

\authorrunning{V. Bürgin et al.}
\institute{ImFusion GmbH, Munich, Germany 
\and
Technical University of Munich, Germany
}
\maketitle              %
\begin{abstract}
Automatic vertebra localization and identification in CT scans is important for numerous clinical applications.
Much progress has been made on this topic, but it mostly targets positional localization of vertebrae, ignoring their orientation.
Additionally, most methods employ heuristics in their pipeline that can be sensitive in real clinical images which tend to contain abnormalities. 
We introduce a simple pipeline that employs a standard prediction with a U-Net, followed by a single graph neural network to associate and classify vertebrae with full orientation.
To test our method, we introduce a new vertebra dataset that also contains pedicle detections that are associated with vertebra bodies, creating a more challenging landmark prediction, association and classification task.
Our method is able to accurately associate the correct body and pedicle landmarks, ignore false positives and classify vertebrae in a simple, fully trainable pipeline avoiding application-specific heuristics.
We show our method outperforms traditional approaches such as Hungarian Matching and Hidden Markov Models.
We also show competitive performance on the standard VerSe challenge body identification task.

\end{abstract}
\section{Introduction}
\begin{figure}[t]
  \centering
  \includegraphics[width=0.95\linewidth]{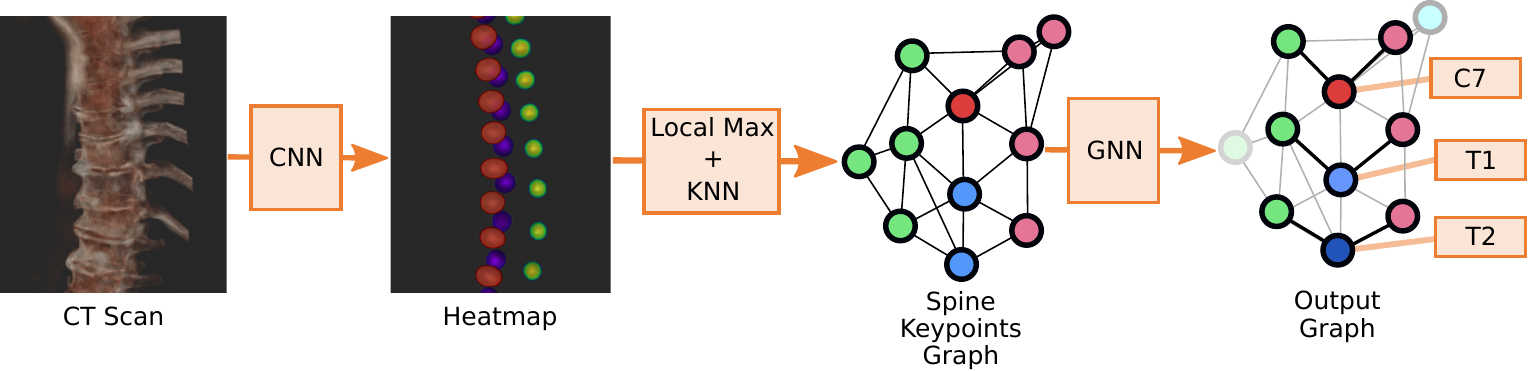}
  \caption{Summary of the proposed method. A CNN generates a heatmap from which local maxima are connected using k-nearest neighbours to form a graph. Then a single GNN associates the keypoints, performs classification and filters out false postives. The full pipeline is trained and does not require hand-tuned post-processing.}
  \label{fig:pipeline}
\end{figure}

Vertebra localization and identification from CT scans is an essential step in medical applications, such as pathology diagnosis, surgical planning, and outcome assessment~\cite{burns2016automated,bourgeois2015evolution}.
This is however a tedious manual task in a time-sensitive setting that can benefit a lot from automation.
However, automatically identifying and determining the location and orientation of each vertebra from a CT scan can be very challenging:
(i) %
scans vary greatly in intensity and constrast,
(ii) %
metal implants and other materials can affect the scan quality,
(iii) %
vertebrae might be deformed, crushed or merged together due to medical conditions,
(iv) %
vertebrae might be missing due to accidents or previous surgical operations.

Recently, public challenges like the VerSe challenge~\cite{sekuboyina2021verse} have offered a common benchmarking platform to evaluate algorithms to automate this task, resulting in a boost in research on this topic.
However, these challenges focus on finding the position of vertebrae, ignoring the orientation or direction.
Additionally, practically all methods employ manual heuristic methods to identify landmarks and filter out false positives.

In this paper, we introduce a trainable method that performs vertebrae localization, orientation estimation and classification with a single architecture.
We replace all hand-crafted rules and post-processing steps with a single trainable Graph Neural Network (GNN) that learns to filter out, associate and classify landmarks.
We apply a generalized Message Passing layer that can perform edge and node classification simultaneously.
This alleviates the need for sensitive hand-tuned parameter tuning, and increases robustness of the overall pipeline.

The main contributions of our work are: (1) introducing a pipeline that uses a single Graph Neural Network to perform simultaneous vertebra identification, landmark association, and false positive pruning, without the need for any heuristic methods and (2) building and releasing a new spine detection dataset that adds pedicles of vertebrae to create a more complex task that includes orientation estimation of vertebrae, which is relevant for clinical applications.

\section{Related Work}
\textbf{Spine Landmark Prediction and Classification} The introduction of standardised spine localization and classification challenges~\cite{glocker2013vertebrae,sekuboyina2021verse} resulted in a boost in research on this problem.
Convolutional Neural Networks (CNN) became a dominant step in most approaches shortly after their introduction~\cite{chen2015automatic}.
Most modern methods process the results of a CNN using a heuristic method to create a 1-dimensional sequence of vertebra detections, before applying classification:
\cite{yang2017automatic} generate heatmaps for body positions, and refine it into a single sequence graph that uses message passing for classification.
\cite{liao2018joint} generate heatmaps, extract a 1-dimensional sequence and use a recurrent neural network for classification.
\cite{payer2020coarse} produce a heatmap using a U-Net~\cite{ronneberger2015u}, but use a simpler approach by taking the local maxima as landmarks, and forming a sequence by accepting the closest vertebra that is within a defined distance range.
\cite{meng2022vertebrae} uses a directional graph and Dynamic Programming to find an optimal classification.

\textbf{Graph Neural Networks}
In recent years, Graph Neural Networks (GNN) have surged in popularity~\cite{gori2005new}, with a wide and growing range of applications~\cite{wu2020comprehensive}. 
A prominent task in the literature is node-level representation learning and classification. 
A less prominent task is edge classification, for which early work used a dual representation to turn edge- into node representations~\cite{bandyopadhyay2019beyond}. 
Other approaches model edge embeddings explicitly, such as \cite{kipf2018node2Edge} and \cite{braso2020multipleObjectTracking}.
The most general formulation of GNNs is the message-passing formulation \cite{bronstein2021geometric} which can be adapted to perform both edge and node classification at the same time.
We use this formulation in our method.

Various methods have applied GNNs to keypoint detection, however they all apply to 2-dimensional input data.
In \cite{reddy2019occlusion} GNNs are used to detect cars in images.
An edge classification task is used to predict occluded parts of the car.
However, the GNN step is ran individually for every car detection and the relation between cars is not taken into account, unlike our task.
Also there is no node classification applied.
In \cite{lin2021learning} a GNN is used to group detected keypoints for human-pose estimation on images.
Keypoints are grouped using edge prediction where edge-embeddings are used as input the the GNN.
A separate GNN processes node embeddings to facilitate the final grouping.
However, the node and edge embeddings are processed separately from each other.

\section{Method}
In this paper we tackle vertebra localization and classification, but unlike other methods that only focus on detecting the body of the vertebrae, we also detect the pedicles and associate them with their corresponding body.
This allows us to also define the \emph{orientation} of the vertebra defined by the plane passing through the body and pedicles~\footnote{Another common way to define the orientation is using the end-plates of the vertebra body; however end-plates can be irregular and ill-defined in pathological cases}.
To this end we introduce a new dataset that includes pedicles for each vertebra, described in Section~\ref{sec:dataset}, creating a challenging keypoint detection and association task.

Our method consists of a two-stage pipeline shown in Figure~\ref{fig:pipeline}: we detect keypoints from image data using a \emph{CNN stage}, and form a connected graph from these keypoints that is processed by a \emph{GNN stage} to perform simultaneous node and edge classification, tackling classification, body to pedicle association as well as false positive detection with a single trainable pipeline without heuristics.

\textbf{CNN stage} The CNN stage detects candidate body, left pedicle and right pedicle keypoints and provides segment classifications for the body keypoints as either cervical, thoracic, lumbar or sacral. %
We use a UNet$^2$ CNN~\cite{qin2020u2} and select all local maxima with an intensity above a certain threshold $\tau$, in this paper we use $\tau=0.5$.
These keypoints are connected to their $k$ nearest neighbours, forming a graph.
In rare cases this can result in unconnected cliques, in which case the nearest keypoint of each clique is connected to $\tfrac{k}{3}$ nearest points in the other cliques, ensuring a fully connected graph.
All nodes and edges are associated with information through embeddings, described below.

\textbf{GNN stage}
The second stage employs a generalized message-passing GNN following \cite{bronstein2021geometric} to perform several prediction tasks on this graph simultaneously:
\begin{enumerate}
    \item \textbf{keypoint association prediction:} we model association between body keypoints and their corresponding pedicle keypoints as binary edge classification on the over-connected $k$-NN graph.
    \item \textbf{body keypoint level prediction:} for body keypoints, we model the spine level prediction as multi-class node classification.
    \item \textbf{keypoint legitimacy prediction:} to filter out false-positive keypoints, we additionally compute an binary legitimacy prediction for each node.
\end{enumerate}
To perform these task, our message-passing GNN maintains edge and node embeddings which are updated in each layer. A message-passing layer performs a node update and edge update operation. Denoting the feature vector of a node $v$ by $x_v$, and the feature vector of a directed edge $(u,v)$ by $x_{uv}$, the node and edge features are updated as follows:

\begin{equation}
    \underbrace{
    x_u' = \bigoplus_{v \in \mathcal N_u \cup \{u\}} \psi_\mathrm{node}(x_u, x_v, x_{uv})}_{\text{Node update}},
    \qquad
    \underbrace{
    x_{uv}' =
            \vphantom{\bigoplus_{v \in \mathcal N_u \cup \{u\}}}
    \psi_\mathrm{edge}(x_u, x_v, x_{uv})
    }_{\text{Edge update}}
\end{equation}

Here $\bigoplus$ denotes a symmetric pooling operation (in our case max pooling) over the neighborhood $\mathcal{N}_u$. 
$\psi_\text{node}$ and $_\text{edge}$ are trainable parametric functions: in our case, two distinct two-layer MLPs with \text{ReLU} nonlinearities. After $N$ such message-passing layers we obtain an embedding vector for each node and edge. Each node/edge embedding is passed through a linear layer (distinct for nodes and edges) to obtain a vector of node class logits or a single edge prediction logit, respectively. The last entry in the node prediction vector is interpreted as a node legitimacy prediction score: nodes predicted as illegitimate are discarded for the output.

The node input features $x_u \in \mathbb R^7$ consist of the one-hot encoded keypoint type (body, left or right pedicle) and the segment input information (a pseudo-probability in $[0, 1]$ for each of the four spine segments of belonging to that segment, computed by applying a sigmoid to the heatmap network's output channels which represent the different spine segments). The edge input features $x_{uv} \in \mathbb R^4$ consist of the normalized direction vector of the edge and the distance between the two endpoints.

The output of the GNN contains finer spine-level classification (i.e. C1-C7, T1-T13, L1-L6, S1-S2), keypoint-level legitimacy (legitimate vs. false-positive detection) and body-pedicle association via edge prediction, implicitly defining the orientation of each vertebra. Prediction scores of corresponding directed edges $(u, v)$ and $(v, u)$ are symmetrized by taking the mean.

In our experiments we consider variations to our architecture: weight sharing between consecutive GNN layers, multiple heads with a shared backbone (jointly trained) and dedicated networks (separately trained) for edge/node prediction.

\section{Experiments}
\label{sec:experiments}

\begin{figure}[t!]
     \centering
    \begin{tabular}{ c | c | c }
     \begin{minipage}[b]{0.16\textwidth}Hungarian Matching\end{minipage}  & \quad \includegraphics[width=4.5cm]{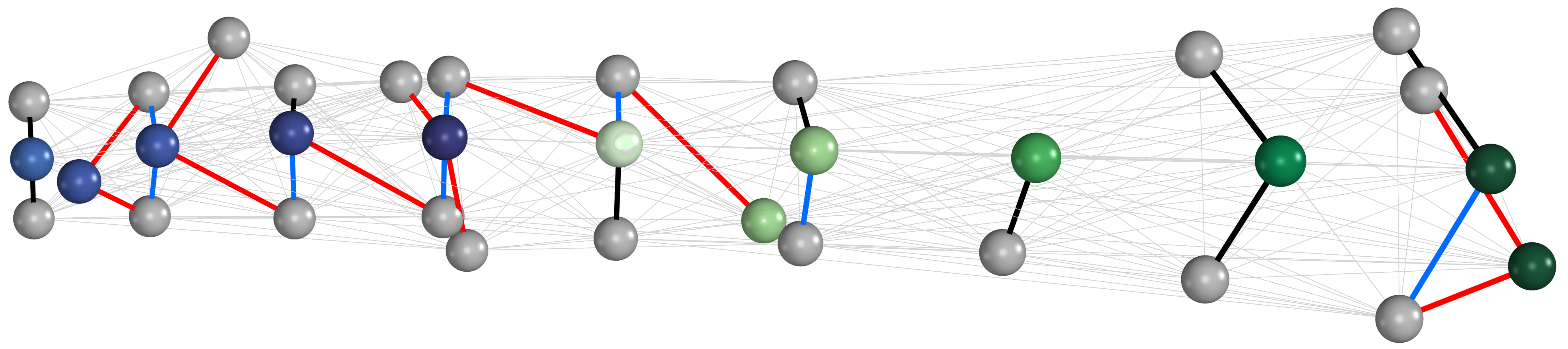} \hspace{0.15cm} & \quad \includegraphics[width=4.5cm]{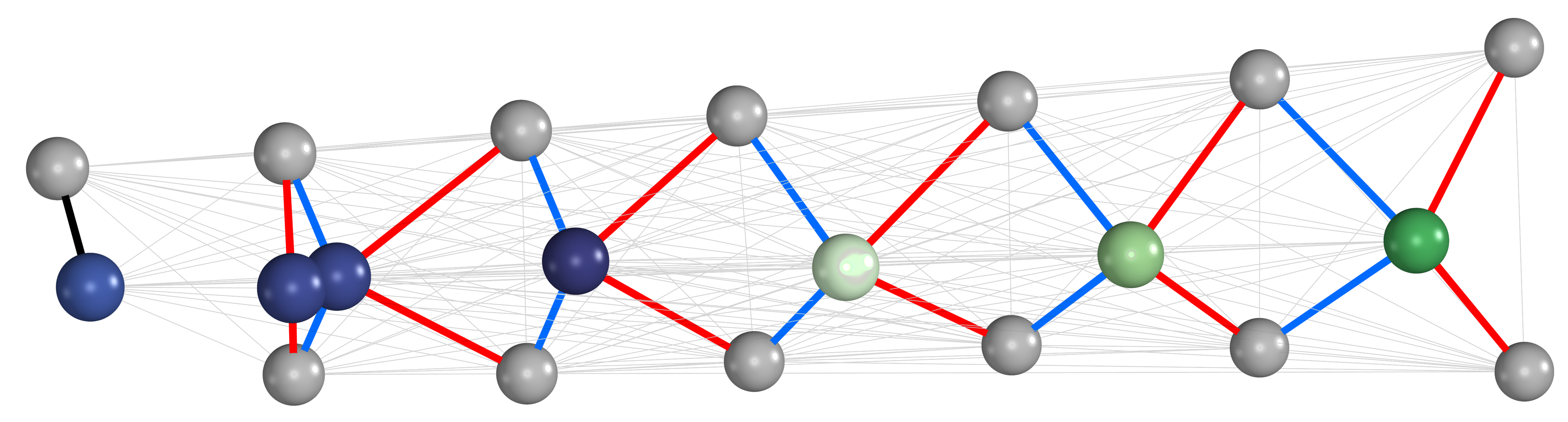} \\ 
     
     \begin{minipage}[b]{0.16\textwidth}GNN \\ (ours)\end{minipage} & \quad \includegraphics[width=4.5cm]{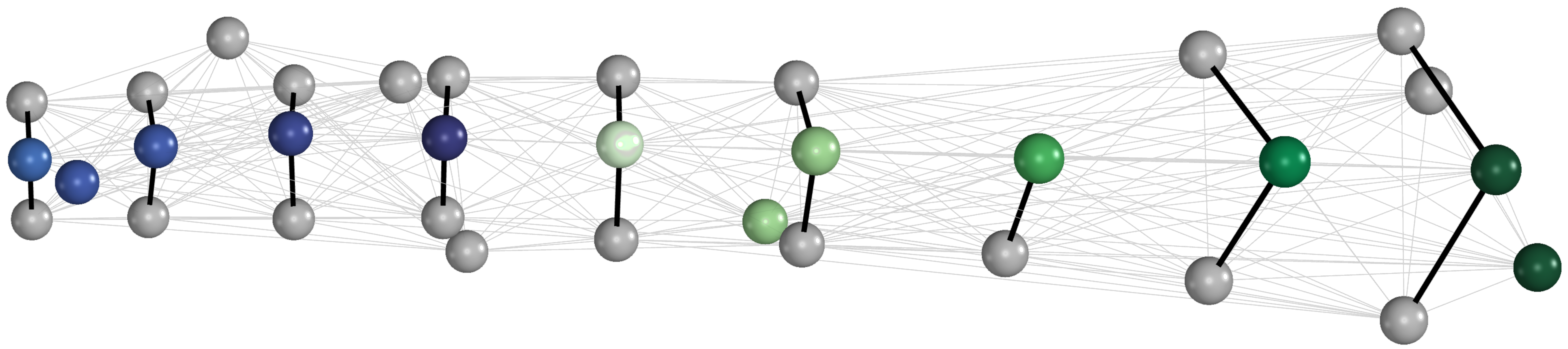} \hspace{0.15cm} & \quad \includegraphics[width=4.5cm]{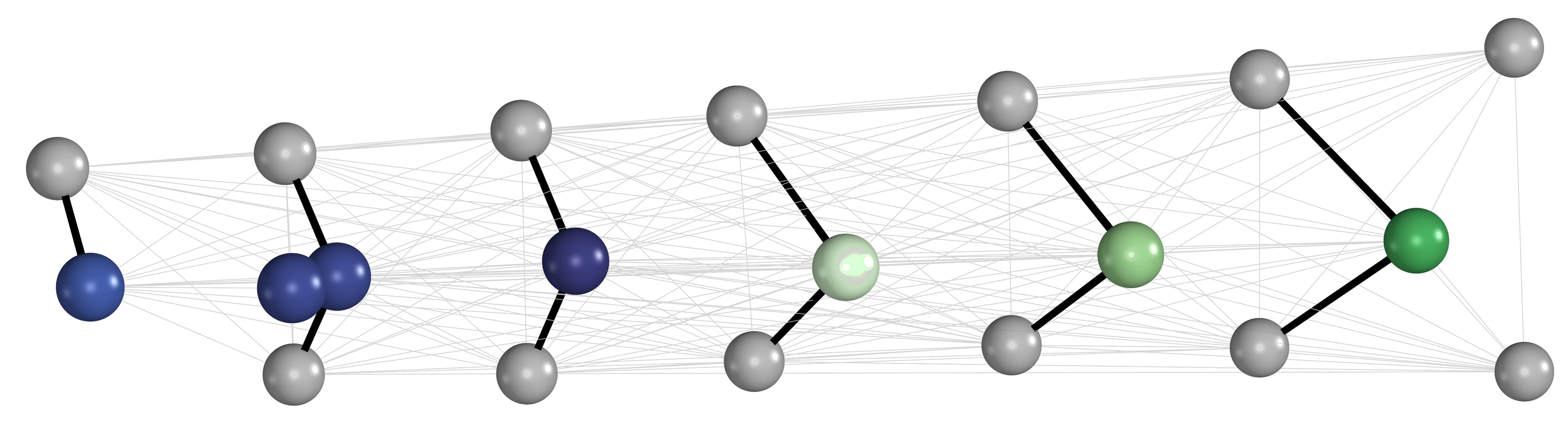} \\   
     \ & Example 1 & Example 2
    \end{tabular}
     \caption{Examples of challenging cases for the Hungarian Matching edge detection baseline, and corresponding correct detections by our GNN architecture (red = false positive, blue = false negative detection).}
     \label{fig:comparison}
\end{figure}

\subsection{Dataset}
\label{sec:dataset}
Our main dataset consists of 2118 scans, of which 1949 form the training dataset, and 169 the validation dataset.
This includes 360 scans from the VerSe datasets \cite{sekuboyina2021verse}, 1676 scans from the CT Colonography dataset~\cite{smith2015data} and 82 scans from the CT Pancreas dataset~\cite{roth2016data}.
Of these datasets, only VerSe has labels for spine levels, and none of the datasets have labels for the pedicles of vertebrae.
The labeling was done in a two-step process with initial pedicle locations determined through post-processing of ground truth segmentations, then transfered to other datasets via bootstrapping.
For each vertebra, the vertebra level is labeled, including the position of the body and the right and left pedicles.
This keypoint dataset is publicly available at \url{https://github.com/ImFusionGmbH/VID-vertebra-identification-dataset}.

Additionally, we also perform the VerSe body classification task on the original VerSe dataset~\cite{sekuboyina2021verse} which contains 160 scans in an 80/40/40 split.

\subsection{Training}
\paragraph{Heatmap network} The heatmap is generated by a UNet$^2$ network \cite{qin2020u2}, with 4 layers and 32 channels. 
The network is trained on crops of $128^3$ voxels with 1.5mm spacing.
The target data consists of Gaussian blobs ($\sigma = 6\text{mm}$) at the landmark positions.
We sample 70\% of the crops around landmark locations, and 30\% randomly from the volume.
Additionally we apply 50\% mirror augmentations, and 20 degree rotations around all axes.
We use the MADGRAD optimizer \cite{defazio2022adaptivity} with a learning rate of $10^{-3}$.
We use a binary cross-entropy loss, with an 80\% weighting towards positive outputs to counter the data's balancing.

\paragraph{Graph Neural Network}
The GNN is implemented in \emph{PyTorch Geometric} 2.0.4 \cite{fey2019fast}.
The three predictions of the graph neural network -- edge classification, node classification and node legitimacy prediction -- are trained via cross-entropy losses which are weighted to obtain the overall loss:
\begin{equation}
    \mathcal L = \alpha \mathcal{L}_\mathrm{edge}^\mathrm{BCE} + \beta \mathcal{L}_\mathrm{node_\mathrm{class}}^\mathrm{CE} + \gamma \mathcal{L}_\mathrm{node_\mathrm{legit}}^\mathrm{BCE}
\end{equation}
We only make edge predictions on edges that run between a body and a pedicle keypoint -- the other edges are only used as propagation edges for the GNN. 
Similarly, we only make spine level predictions on body keypoints. Solely these subsets of nodes/edges go into the respective losses.

As input data, we use the predicted keypoints of the heatmap network on the training/validation set. 
The ground-truth keypoints are associated to this graph to create the target of the GNN. We include three synthetic model spines during training (keypoints in a line spaced 30mm apart) to show the network typical configurations and all potential levels (with all levels/without T13, L6, S2/without T12, T13, L6, S2).

We tune various hyperparameters of our method, such as network depth and weight sharing, the $k$ of the $k$-NN graph, the loss weighting parameters $\alpha, \beta, \gamma$ and the number of hidden channels in the message-passing MLP.
We use a short notation for our architectures, such as (5x1, 4, 1) for 5 independent message-passing layers followed by 4 message-passing layers with shared weights and another independent message-passing layer.

Various data augmentations are used to make our network more robust to overfitting: %
(i) rotation of the spine by small random angles,
(ii) mirroring along the saggital axis (relabeling left/right pedicles to keep consistency),
(iii) perturbation of keypoints by small random distances,
(iv) keypoint duplication and displacement by a small distance (to emulate false-positive duplicate detections of the same keypoint),
(v) keypoint duplication and displacement by a large distance (to emulate false-positive detections in unrelated parts of the scan) and
(vi) random falsification of spine segment input features.
We define four levels of augmentation strength (no/light/default/heavy augmentations) and refer to the supplementary material for precise definitions of these levels.

\subsection{Evaluation}
We evaluate our method on two tasks. The first one is the full task consisting of node and edge classification for vertebra level and keypoint association detection. We evaluate this on the 2118 scan dataset which comes with pedicle annotations.
The second task consists only of vertebra localization, which we evaluate on the VerSe 2019 dataset \cite{sekuboyina2021verse}. Unless otherwise specified, we use $k=14$, the (13x1) architecture, batch size 25 and reaugment every 25 epochs for the full task, and $k=4$, the (9x1) architecture, batch size 1 and reaugment every epoch for the VerSe task. In both cases we use the default augmentation level and $\alpha = \beta = 1$.
As evaluation metrics we use the VerSe metrics \emph{identification rate} (ratio of ground-truth body keypoints for which the closest predicted point is correctly classified and within 20mm) and \emph{$d_\text{mean}$} (mean distance of correctly identified body keypoints to their 1-NN predictions) \cite{sekuboyina2021verse}. Furthermore we evaluate the edge and illegitimacy binary predictions by their $F_1$ scores. Since the identification rate is largely unaffected by false-positive predicted keypoints, we disable legitimacy predictions unless for specific legitimacy prediction experiments to help comparability.

We compare our methods to two baselines: 
For node prediction, we use a Hidden Markov Model (HMM)~\cite{baum1966statistical} that is fitted to the training data using the Baum-Welch algorithm using the \emph{pomegranate} library \cite{schreiber2017pomegranate}. The HMM gets the predicted segment labels in sequence. Dealing with false-positive detection outliers is very difficult for this baseline, therefore we filter out non-legitimate detections for the HMM inputs to get a fairer comparison, making the task slightly easier for the HMM. 
For the VerSe challenge, we also compare our results to the top papers reported in the VerSe challenge~\cite{sekuboyina2021verse}.
For edge prediction, we compare our method to Hungarian matching on the keypoints from the CNN pipeline.

\newcommand{\dummypercent}{12.34\%}
\newcommand{\dummypm}{0.99}

\begin{table}[!t]
    \caption{\footnotesize Results on the 2118 spine dataset (full validation set / hard subset). Comparing best GNN architectures with the baselines. Wilcoxon signed-rank test $p$-values are given for GNN numbers that outperform the baseline (by the test's construction, $p$-values agree between the full and hard subset).
    }
    \vspace{0.7ex}    
    \setlength{\tabcolsep}{7pt}
    \renewcommand{\arraystretch}{1.2}
    \centering
    \resizebox{0.99\textwidth}{!}{
    \begin{tabular}{c c | c c c}
    \toprule
    \multicolumn{2}{c |}{Method} & identification rate & edge $F_1$ score & $d_\mathrm{mean}$
    \\
    \midrule
    \multirow{4}{*}{\shortstack{Edge vs. node\\classification} \vspace{0.5cm}} 
        & \shortstack{Joint node/edge\\~} & \shortstack{\textbf{97.19}\,/\,\textbf{89.88}\\[-0.9ex]\scriptsize($p=0.080$)} & \shortstack{98.81\,/\,96.22\\[-0.9ex]\vphantom{\scriptsize($p$)}} & \shortstack{\textbf{1.68}\,/\,\textbf{1.95}\\[-0.9ex]\vphantom{\scriptsize($p$)}}
        \\
        & \shortstack{Node only\\~} & \shortstack{96.91\,/\,88.90\\[-0.9ex]\scriptsize($p=0.109$)} & \shortstack{\textbf{\textemdash}\\[-0.9ex]\vphantom{\scriptsize($p$)}} & \shortstack{1.68\,/\,1.96\\[-0.9ex]\vphantom{\scriptsize($p$)}}
        \\
        & \shortstack{Edge only\\~} & \shortstack{\textbf{\textemdash}\\[-0.9ex]\vphantom{\scriptsize($p$)}} & \shortstack{\textbf{99.31}\,/\,\textbf{97.77}\\[-0.9ex]\scriptsize($p=0.019$)} & \shortstack{\textbf{\textemdash}\\[-0.9ex]\vphantom{\scriptsize($p$)}}
        \\  
    \midrule
    \multirow{4}{*}{\vspace{0.9cm}\shortstack{Baselines}}
        & Hidden Markov & 94.29\,/\,79.46 & \textbf{\textemdash}  & 1.81\,/\,2.52
        \\
        & Hungarian matching & \textbf{\textemdash} & 98.93\,/\,96.56 & \textbf{\textemdash}
        \\
    \bottomrule
    \end{tabular}
    } \label{tab:2118results}
\end{table}

\begin{table}[!htb]
    \setlength{\tabcolsep}{7pt}
    \renewcommand{\arraystretch}{1.2}
    \centering
    \caption{\footnotesize Hyperparameter comparisons on the 2118 spine dataset. Comparing GNN architectures, influence of enabling legitimacy predictions and of different augmentation strengths. Best values within each group highlighted in bold.
    }
    \vspace{0.7ex}
    \resizebox{0.99\textwidth}{!}{
    \begin{tabular}{c c | c c c}
    \toprule
    \multicolumn{2}{c |}{Method} & identification rate & edge $F_1$ score
    & illegitimacy $F_1$
    \\
    \midrule
    \multirow{4}{*}{\shortstack{Single-head\\architectures}\vspace{0cm}}
        & (7x1) & 96.74 & 98.86 
        & \textbf{\textemdash} 
        \\
        & (13x1) & \textbf{97.19} & 98.81 
        & \textbf{\textemdash} 
        \\
        & (1,5,1) & 96.75 & 98.91 
        & \textbf{\textemdash} 
        \\
        & (1,11,1) & 97.10 & 98.67
        & \textbf{\textemdash} 
        \\
    \midrule
    \multirow{4}{*}{\vspace{0cm}\shortstack{Multi-head\\architectures:\\backbone,\\edge/node head}\vspace{0.4cm}}
        & (1x1), (4x1), (12x1) & 96.94 & 99.16 
        & \textbf{\textemdash} 
        \\
        & (3x1), (2x1), (10x1) & 96.87 & 99.00 
        & \textbf{\textemdash} 
        \\
        & (5x1), (\textbf{\textemdash}) , (8x1) & 96.79 & 98.88 
        & \textbf{\textemdash} 
        \\
    \midrule
    \multirow{4}{*}{\vspace{0.5cm}\shortstack{Dedicated edge\\architectures}}
        & (3x1) & \textbf{\textemdash} & 99.28 
        & \textbf{\textemdash} 
        \\
        & (5x1) & \textbf{\textemdash} & 99.28 
        & \textbf{\textemdash} 
        \\
        & (1,3,1) & \textbf{\textemdash} & \textbf{99.35}
        & \textbf{\textemdash} 
        \\
    \specialrule{.14em}{.12em}{.12em}
    \multirow{4}{*}{\vspace{0cm}\shortstack{With legitimacy\\prediction: different\\legit. loss weights}}
        & $\gamma = 0.1$ & 95.61 & \textbf{98.96} 
        & 56.00
        \\
        & $\gamma =  1.0$ & \textbf{96.50} & 98.75 
        & 62.41
        \\
        & $\gamma = 5.0$ & 96.35 & 98.18 
        & 61.67
        \\
        & $\gamma = 10.0$ & 96.42 & 98.96 
        & \textbf{62.78}
        \\
    \specialrule{.14em}{.12em}{.12em}
    \multirow{4}{*}{\vspace{0cm}\shortstack{Augmentations}}
        & None & 95.27 & 96.83 
        & \textbf{\textemdash} 
        \\
        & Light & 97.18 & \textbf{98.83} 
        & \textbf{\textemdash} 
        \\
        & Default & \textbf{97.19} & 98.81 
        & \textbf{\textemdash} 
        \\
        & Heavy & 97.03 & 98.67 
        & \textbf{\textemdash} 
        \\
    \bottomrule
    \end{tabular}
    }
    \label{tab:architecturesearch}

    \caption{\footnotesize Results on the VerSe 2019 dataset (validation/test set). We compare our GNN architecture, our Hidden Markov baseline, and the reported numbers of the three top VerSe challenge entries \cite{sekuboyina2021verse}. Best values highlighted in bold.
    }
    \vspace{0.7ex}
    \resizebox{0.99\textwidth}{!}{
    \begin{tabular}{c c | r r r}
    \toprule
    \multicolumn{2}{c |}{Method} & identification rate & \multicolumn{1}{c}{$d_\mathrm{mean}$} & illegitimacy $F_1$
    \\
    \midrule
    \shortstack{Single-head GNN\\[-0.9ex] \vphantom{\scriptsize($p$)}}
        & \shortstack{(9x1)\\[-0.9ex] \vphantom{\scriptsize($p$)}} & \shortstack{93.26\,/\,93.02 \\[-0.9ex] {\scriptsize($p{=}3.1\mathrm{e}{\shortminus6}$/$p{=}5.2\mathrm{e}{\shortminus7}$)}} & \shortstack{1.28\,/\,1.43 \\[-0.9ex] \vphantom{\scriptsize($p$)}} & \textbf{\textemdash} 
        \\
    \midrule
    \shortstack{With legitimacy\\prediction}
        & \shortstack{$\gamma = 10.0$\\[-0.9ex]\vphantom{\scriptsize($p$)}} & \shortstack{90.75\,/\,87.51\\[-0.9ex]{\scriptsize($p{=}4.8\mathrm{e}{\shortminus6}$/$p{=}9.4\mathrm{e}{\shortminus7}$)}} & \shortstack{\textbf{1.23}\,/\,\textbf{1.32}\\[-0.9ex]\vphantom{\scriptsize($p$)}} & \shortstack{77.67\,/\,81.69\\[-0.9ex]\vphantom{\scriptsize($p$)}}
        \\
    \midrule
    \multirow{1}{*}{\vspace{0.9cm}\shortstack{Baseline}}
        & Hidden Markov & 48.59\,/\,49.06 & 1.32\,/\,1.45 & \textbf{\textemdash} 
        \\
    \midrule
    \multirow{4}{*}{\vspace{3ex}\shortstack{VerSe challenge\\entries \cite{sekuboyina2021verse}}}
        & Payer C. \cite{payer2019vertebrae} & 95.65\,/\,\textbf{94.25}  & 4.27\,/\,4.80 & \textbf{\textemdash} 
        \\
        & Lessmann N. \cite{lessmann2019iterative} & 89.86\,/\,90.42 & 14.12\,/\,7.04 & \textbf{\textemdash} 
        \\
        & Chen M. \cite{sekuboyina2021verse} & \textbf{96.94}\,/\,86.73 & 4.43\,/\,7.13 & \textbf{\textemdash} 
        \\
    \bottomrule
    \end{tabular}
    }
    \label{tab:verseresults}
\end{table}

\subsection{Results}
The results on our introduced dataset are shown in Table~\ref{tab:2118results}. Our method outperforms the baseline methods, both in identification rate and edge accuracy. We also show results on a 'hard' subset, selected as the data samples where either of the methods disagree with the ground truth. This subset contains 47 scans and represents harder cases where the association is not trivial. We give $p$-values of the Wilcoxon signed-ranked test against the respective baseline for numbers that are better than the baseline. Figure~\ref{fig:comparison} shows a qualitative example where the baseline method fails due to false-positive and false-negative misdetections in the input (these errors in the input were generated by augmentations and are more extreme than usual, to demonstrate several typical baseline failures in one example). The GNN correctly learns to find the correct association and is not derailed by the misdetections.

The examples where our architecture fails typically have off-by-one errors in the output from the CNN: for example, the last thoracic vertebra is detected as a lumbar vertebra (usually in edge cases where the two types are hard to distinguish). Hence the GNN classifications of the lumbar segment, and possibly the thoracic segment, will be off by one (see Figure S2 in the supplementary).

Table~\ref{tab:architecturesearch} shows the performance differences of various GNN architectures. The single-head architecture with 13 individual layers performs the best on identification rate, although the 13-layer architecture with 11 shared layers performs very similarly.
The architectures with fewer layers perform slightly better in edge accuracy since this task is less context dependent.
Multi-head architectures perform slightly worse on identification, but retain a good edge accuracy.
Training a model solely directed to edge classification does yield the highest edge accuracy, as expected.
Enabling legitimacy predictions slightly degrades the performance, likely due to two facts: for one, an extra loss term is added, which makes the model harder to train. Also, the identification rate metric is not majorly affected by having additional false-positive detections, hence there is little to be gained in terms of this metric by filtering out false positives.
An optimal legitimacy loss weighting seems to be $\lambda = 1.0$.
Finally, augmentations help generalization of the model, but the amount of augmentations seems to have little effect.

Table~\ref{tab:verseresults} shows the result on the traditional VerSe body-identification task.
Our method yields a competitive performance despite not being optimized on this task (identification rate slightly lower than the leaders but a better average distance to the landmarks).

\section{Conclusion}
We introduced a simple pipeline consisting of a CNN followed by a single GNN to perform complex vertebra localization, identification and keypoint association.
We introduced a new more complex vertebra detection dataset that includes associated pedicles defining the full orientation of each vertebra, to test our method.
We show that our method can learn to associate and classify correctly with a single GNN that performs simultaneous edge and node classification.

The method is fully trainable and avoids most heuristics of other methods.
We also show competitive performance on the VerSe body-identification dataset, a dataset the method was not optimized for.
We believe this method is general enough to be usable for many other detection and association tasks, which we will explore in the future.
\FloatBarrier

\bibliographystyle{splncs04}
\bibliography{paper}

\newpage
\input{supplementary}

\end{document}

%% file: supplementary.tex
\section*{Supplementary}

\setcounter{table}{0}
\setcounter{figure}{0}
\renewcommand{\thetable}{S\arabic{table}}
\renewcommand{\thefigure}{S\arabic{figure}}

\begin{table}[h!]
    \setlength{\tabcolsep}{7pt}
    \renewcommand{\arraystretch}{1.2}
    \centering
    \caption{Runtime and memory requirements for the two parts of our method (CNN/GNN part).
    For inference time we give estimates for small/large instances in the respective dataset. All experiments were done on a system with four \textsl{\textsc{Nvidia GTX 1080 TI}} GPUs and an \emph{\textsc{Intel Xeon E5-2680 v4 (2.40GHz)}} CPU, where the CNN training used all four GPUs simultaneously and the GNN training used only one GPU.
    VerSe2019 GNN training time is higher because of more frequent re-augmentation.
    \vspace{0.3cm}
    }
    \resizebox{0.99\textwidth}{!}{
    \begin{tabular}{c | c r r}
    \toprule
    \shortstack{Method and Dataset} & \shortstack{Training time} & \multicolumn{1}{c}{Inference time} & \multicolumn{1}{c}{Memory footprint (training)}
    \\
    \midrule
    \multirow{1}{*}{CNN (2118 scan dataset)}
        & \multirow{1}{*}{$\approx$ 13h} & $\approx$ \phantom{0}5 - 15s\phantom{m} & $\approx$ 4 $\times$ 10GB (GPU)\phantom{.}
        \\
    \midrule
    \multirow{1}{*}{CNN (Verse2019 dataset)}
        & \multirow{1}{*}{$\approx$ \phantom{0}8h} & $\approx$ \phantom{0}5 - 15s\phantom{m} & $\approx$ 4 $\times$ 10GB (GPU)\phantom{.}
        \\
    \midrule
    \multirow{2}{*}{GNN (2118 scan dataset)} 
        & \multirow{2}{*}{2h 12min} & GPU: $\approx$ 13 - 15ms & \multirow{2}{*}{$\approx$ 2.1GB (GPU)\phantom{ $\times$ 2}}
        \\
        & & CPU: $\approx$ 10 - 40ms &
        \\
    \midrule
    \multirow{2}{*}{GNN (Verse2019 dataset)}
        & \multirow{2}{*}{2h 21min} & GPU: $\approx$ \phantom{0}9 - \phantom{0}9ms & \multirow{2}{*}{$\approx$ 0.9GB (GPU)}\phantom{ $\times$ 2}
        \\
        & & CPU: $\approx$ 10 - 30ms &
        \\  
    \bottomrule
    \end{tabular}
    }
    \label{tab:runtimeMemoryRequirements}
\end{table}

\begin{table}[h!]
    \setlength{\tabcolsep}{7pt}
    \renewcommand{\arraystretch}{1.2}
    \centering
    \caption{Probabilities (\%) and parameter values for each of our augmentations at different augmentation levels (light/default/heavy). Each augmentation is applied independently to each node with the given probability, except for the ones marked with ${}^\dagger$ which are applied to the whole graph with the given probability. Augmentations are applied before adding $k$-NN edges to the graph, and in the order as specified here. 
    \vspace{0.3cm}
    }
    \resizebox{0.99\textwidth}{!}{
    \begin{tabular}{l | c c c}
    \toprule
    Augmentation & Light & Default & Heavy
    \\
    \midrule
    falsify segment input label to random other label
    & 0.5 & 1.0 & 2.0 \\
    delete body / pedicle keypoint
    & 0.5 / 2.0 & 2.0 / 5.0 & 7.5 / 15.0 \\
    clone and displace body / pedicle keypoint by $d \sim \mathcal U_{[0.5, 3]}$ cm
    & 5.0 / 5.0 & 10.0 / 10.0 & 15.0 / 15.0 \\
    clone and displace body / pedicle keypoint by $d \sim \mathcal U_{[20, 50]}$ cm
    & 5.0 / 5.0 & 10.0 / 10.0 & 15.0 / 15.0 \\
    ${}^\dagger$ mirror along saggital ($x$) axis
    & 50.0 & 50.0 & 50.0 \\ 
    ${}^\dagger$ scale along $x$/$z$ axis by $c_x \sim \mathcal U_{[0.8, 1.2]}$ / $c_z \sim \mathcal U_{[0.5, 1.5]}$
    & 5.0 & 10.0 & 30.0 \\ 
    ${}^\dagger$  rotate along $z$/$y$/$x$ axis by $\theta_z, \theta_y \sim \mathcal U_{[-20, 20]}$\textdegree / $\theta_x \sim \mathcal U_{[-40, 40]}$\textdegree
    & 5.0 & 10.0 & 30.0 \\
    perturb by $v \sim \mathcal N(0, (\tfrac{d}{3})^2 \cdot I)$, $|v|$ truncated to $d$ cm (prob.; $d$)
    & 20.0; 0.1cm & 50.0; 0.2cm & 50.0; 0.4cm \\
    \bottomrule
    \end{tabular}
    }
    \label{tab:augmentations}
\end{table}

\begin{figure}[t]
  \centering
  \includegraphics[width=0.95\linewidth]{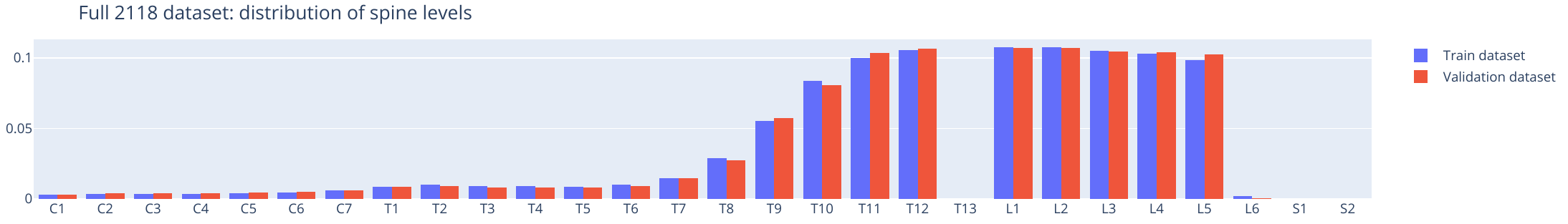}
  \vspace{0.5cm}
  \includegraphics[width=0.95\linewidth]{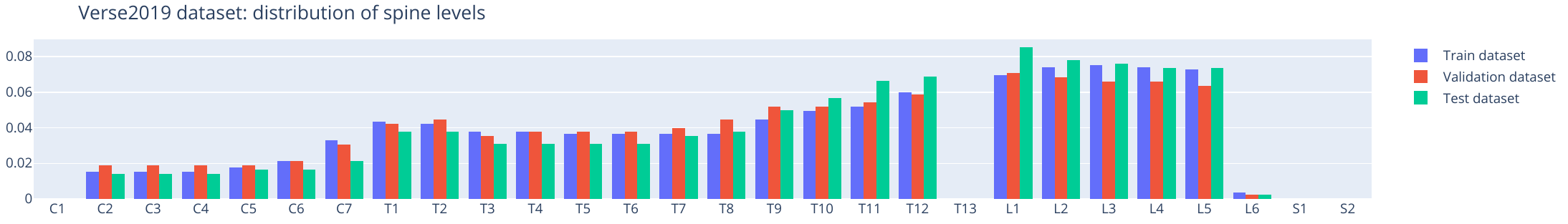}
  \caption{Distributions of individual spine levels in our two datasets.}
  \label{fig:levelsDistribution}
\end{figure}

\begin{figure}
    \centering
    \begin{subfigure}[b]{0.49\textwidth}
        \centering
        \includegraphics[width=4.9cm]{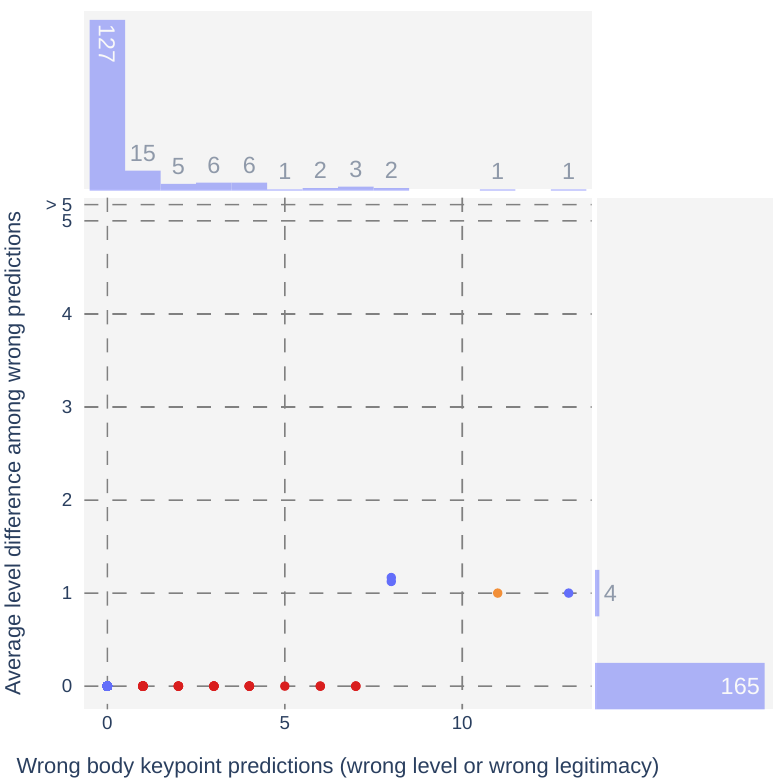}
        \caption{2118 scan dataset, GNN (13x1), $k=14$}
    \end{subfigure}
    \hfill
    \begin{subfigure}[b]{0.49\textwidth}
        \centering
        \includegraphics[width=4.9cm]{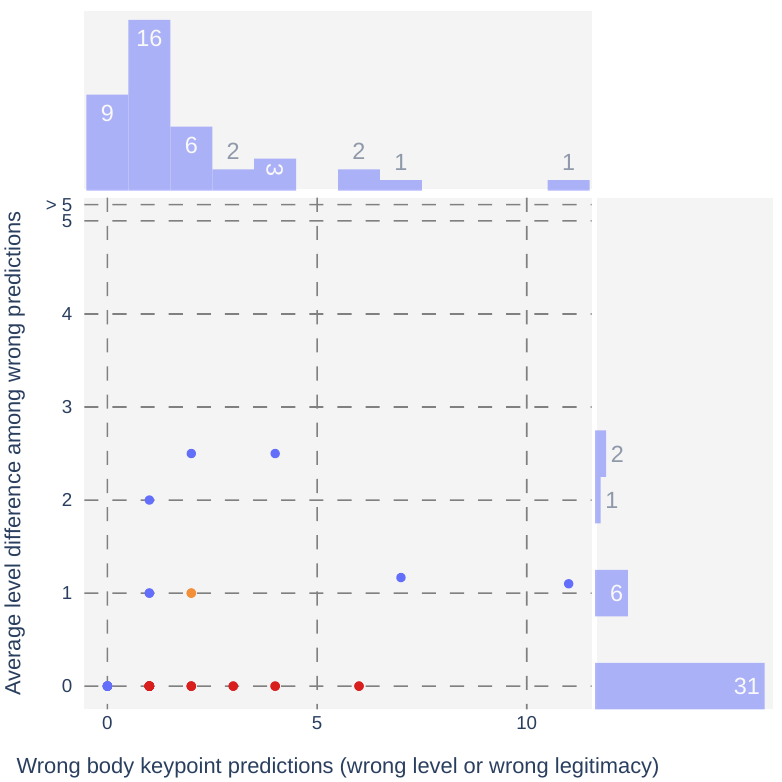}
        \caption{Verse2019 dataset, GNN (9x1), $k=4$}
    \end{subfigure}
     \begin{subfigure}[b]{0.49\textwidth}
         \centering
         \includegraphics[width=4.9cm]{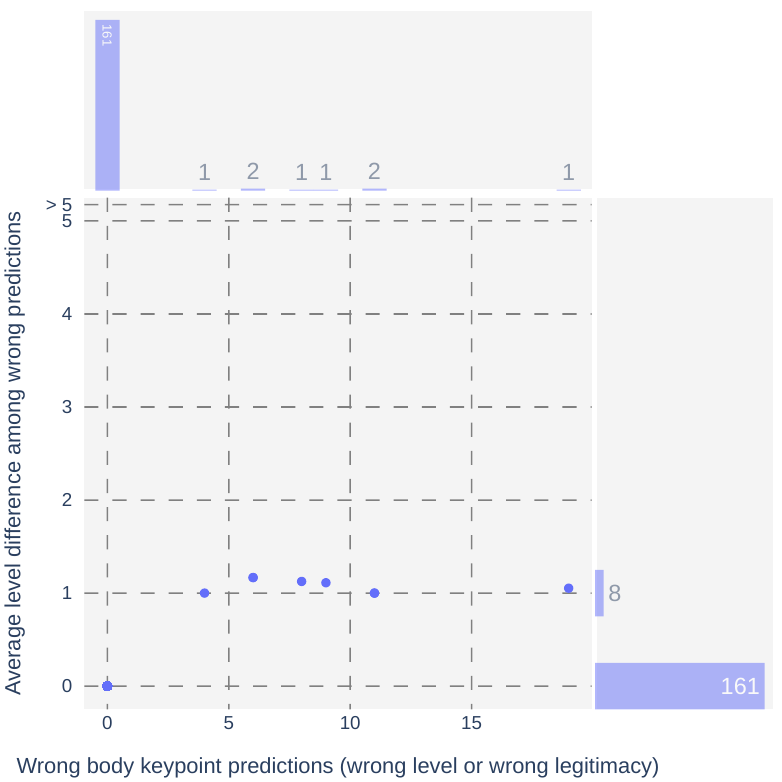}
         \caption{2118 scan dataset, HMM baseline}
     \end{subfigure}
     \hfill
     \begin{subfigure}[b]{0.49\textwidth}
         \centering
         \includegraphics[width=4.9cm]{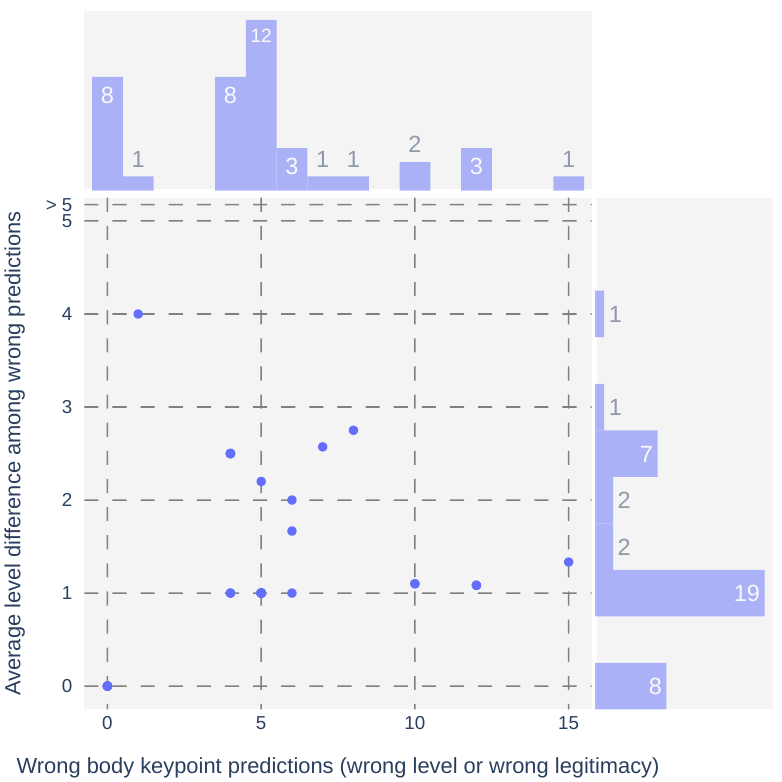}
         \caption{Verse2019 dataset, HMM baseline}
     \end{subfigure}
     
    \caption{Scatterplots of wrongly predicted levels vs. average level difference on the different datasets/architectures. Each dot is one spine, the number of wrongly predicted levels on the $x$ axis and the average level difference for wrong predictions (legitimate nodes only) on the $y$ axis ($y=1$: all errors are off-by-one errors). Redder color means more false-positive legitimacy predictions. The red points at $y=0$ are spines where the only errors are false-positive legitimacy predictions (but the levels of all legitimate keypoints are correctly identified).}
    \label{fig:scatterplots}
\end{figure}